\documentclass[useAMS,usenatbib]{mn2e}
\usepackage{graphicx}
\usepackage{psfig}
\usepackage{amsfonts,amssymb,amsmath}
\usepackage{hyperref,color}
\usepackage{textcomp}
\usepackage[T1]{fontenc}
\usepackage{aecompl}
\usepackage{times}
\usepackage[colorinlistoftodos]{todonotes}
\usepackage{txfonts}
\newcommand{\aap}    {A\&A}
\newcommand{\apjs}   {ApJS}

\newcommand{\apjl}   {ApJL}

\title[Merging Systems Identification]{The MeSsI (Merging Systems Identification) Algorithm \& Catalogue.}
\author[Mart\'{\i}n de los Rios, Mariano J. Dom\'{\i}nguez R., Dante Paz, Manuel Merch\'an.]
{Mart\'{\i}n de los Rios\thanks{E-mail:martin@oac.unc.edu.ar, mardom@oac.uncor.edu},
Mariano J. Dom\'{\i}nguez R.\footnotemark[1], Dante Paz, Manuel Merch\'an $^{1,2,3}$.\\
$^{1}$ Instituto de Astronom\'{\i}a Te\'orica y Experimental (CCT C\'ordoba, CON
ICET, UNC), Laprida 854, X5000BGR, C\'ordoba, Argentina.\\
$^{2}$ Observatorio Astron\'omico de C\'ordoba, Universidad Nacional de C\'ordoba, Laprida 854, X5000BGR, C\'ordoba, Argentina.\\
$^{3}$ Consejo Nacional de Investigaciones Cient\'{\i}ficas y T\'ecnicas, Rivadavia 1917, C1033AAJ Buenos Aires, Argentina.}

\begin{document}

\date{Accepted XXX Received XXX; in original form XXX}

\pagerange{\pageref{firstpage}--\pageref{lastpage}} \pubyear{2015}

\maketitle
 
\label{firstpage}

\begin{abstract}
   Merging galaxy systems provide observational evidence
of the existence of dark matter and constraints on its properties. Therefore,
statisticaly uniform samples of merging systems would be a powerful tool for several studies.

In this work we present a new methodology for the identification of merging systems and the results of its application to galaxy redshift surveys.

We use as a starting point a mock catalogue of galaxy systems, identified using FoF algorithms, which experienced a major merger as indicated by its merger tree.

Applying machine learning techniques in this training sample, and using several features computed from the 
 observable properties of galaxy members, it is possible to select galaxy groups with a high probability of having experienced a major merger. Next we apply a mixture of Gaussian technique on galaxy members in order to reconstruct the properties of the haloes involved in such merger.
This methodology provides a highly reliable sample of merging systems with low contamination and precisely recovered properties.
We apply our techniques to samples of galaxy systems obtained from SDSS-DR7, WINGS and HeCS.
Our results recover previously known merging systems and provide several new candidates. We present their measured properties and discuss future analysis on current and forthcoming samples.  

\end{abstract}


\section{Introduction.}

Merging galaxy systems such as the Bullet Cluster \citep{clowe}, Abell 520 \citep{mahdavi,jee2012,clowe2012,jee2014}, Baby Bullet \citep{bradac}, Pandora \citep{merten}, Musket Ball \citep{dawson}, El Gordo \citep{menanteau,dawsongordo,molnar}, Abell 1758 \citep{durret} and Abell 3716 \citep{andrade} have provided observational evidence for the existence of dark matter.
 Most of them have been used to test the Cold Dark Matter (CDM) paradigm itself (\cite{markevitch}, \cite{hayashi}, \cite{farrar},
 \cite{milosav}, \cite{springel}, \cite{randall}, \cite{mastro}, \cite{lee2}, \cite{forero},
\cite{thompson}, \cite{watson}, \cite{challenger}). 
Several statistical techniques have been proposed to measure dark matter properties using merging systems \citep[such as the self interaction
cross section,][]{massey,harvey,kahlhoefer,harveytwo}. However, the lack of a complete and
 uniformly selected sample of merging systems prevents efforts to derive robust constraints.
In order to overcome this limitation different approaches have been proposed: the \href{www.mergingclustercollaboration.org}{Merging Cluster Collaboration} uses the radio emission due to induced shocks in the Intra Cluster Medium \citep[ICM,][]{feretti} to obtain high redshift merging system candidates.
These systems have been studied using pan-chromatic observations and detailed merging kinematic Bayesian reconstructions \citep{2013ApJ...772..131D}. 
X-ray imaging, spectra and Sunyaev Zeldovich effect observations have been used to identify unrelaxed clusters of galaxies \citep{mann}, cluster mergers \citep{harvey}, substructures and any departures from hydrostatic equilibrium, mainly for the most massive galaxy clusters.

Galaxy redshift surveys are very useful to trace the dynamical state of galaxy systems and to search for substructures \citep{dressler}.
To this end, some methods look for departures in the global Gaussian redshift distribution of system members
 \citep{solanes,hou,serra}. Even though all these methods aim to identify the 
 substructures and recover its properties, they suffer from false identifications and incompleteness, at least to some extent.\\
In this work we develop a uniform identification algorithm of merging systems based on galaxy redshift catalogues. 
These methods can be applied on low mass systems and should increase the
 number of merging systems like the bullet group recently identified by \cite{gael}.
This paper is organized as follows: In Sec.\ref{sec:method}, we apply machine learning techniques to a number of observable features and
 present its calibration based on the result of simulations. We also introduce techniques to recover properties of merging dark matter haloes. 
In Sec.\,\ref{sec:samples} we apply our techniques to samples of galaxy systems identified from low-redshift galaxy surveys
 (\href{http://classic.sdss.org/dr7/}{SDSS-DR7}, HeCS, WINGS). 

 Finally, in Sec.\,\ref{sec:conclusions} we summarize the main results of this work and discuss uses of this new sample
 of merging systems. We adopt the standard cosmological model used in the Millenium Simulation \citep{springel:05} when necessary
  ($H_{0}$\,=\,73\,km\,s$^{-1}$\,Mpc$^{-1}$, $ \Omega_{m} $\,=\,0.25, and $ \Omega_{\Lambda} $\,=\,0.75).


\section[]{Methodology.}
\label{sec:method}

\subsection{Mock Galaxy and Halo Catalogues}
From the point  of view of the current theory of galaxy  formation, the most direct route for defining galaxy systems is 
via its host dark matter halo
\citep{Mo_Van_den_Bosch_White_2010}. 

The Millennium simulation \citep{springel:05}, used in this work, provides a catalogue of dark matter (sub)haloes constructed using a traditional 3D Friends-of-Friends algorithm (FoF) that percolates nearby particles.
The \href{http://www.mpa-garching.mpg.de/millennium/}{GAVO Millennium data base}, 
also provides merger trees for each halo \citep{roukema}, therefore given all the haloes belonging to a FoF group, it is possible to infer its merger tree.
We compute for each FoF group a parent list of FoF progenitors
(identified in previous snapshots) which have contributed with at least one subhalo to the current FoF group. 
We define a major merger of FoF groups as the merger between two groups where the total mass of the involved 
haloes 
represents at least a 20 per cent of the mass after the merger, it is worth noting that this condition imposes a minimal value of 0.25 for the mass ratio of the interacting systems.
We construct mock catalogues of the SDSS-DR7 redshift survey based on the results of a semi-analytic model \citep{guo} and use it to calibrate our 
identification method for merging systems. This process is extensively described in previous works \citep{lares} and \citep{dominguez}. 

We define a recent major merger as a the major merger of two FoF groups,
where its principal haloes are still present as different haloes in the final FoF group.
With this selection criteria the mean of the look-back merger time is around 3 Gyr consistent with other work \citep{pinkney}.
The FoF groups identified as recent major mergers and their member galaxies identified in the mock catalogs
are used to train different machine learning methods.

\subsection{Identification Technique of Merging Systems.} 
Galaxy systems were identified in the mock catalogue reproducing the FoF process as applied on real redshift catalogues (\cite{manuel},
 \cite{ariel}). 

Using the properties of the galaxies in these FoF systems we compute several features relevant to the problem, namely:
\begin{enumerate}

 \item The DS test developed by \cite{dressler}, uses the deviation of the local radial velocity, defined as the mean radial velocity of the closest $n$ galaxies to each galaxy, from the global radial velocity in order to find substructures in clusters of galaxies.
 A global cluster value of the DS test is then obtained by summing up individual galaxy values. Following \cite{pinkney} we select those systems of galaxies with an occupancy $Ngal > 30$ galaxy members in order to have a better identification.
 We performed the DS test for $n=10$ and $n=\sqrt{Ngal}$ and use global and individual DS values as features.

 \item Well known tests measuring the departures from a normal Gaussian distribution: The Anderson-Darlling test, the Cramer-von Mises test, Kolmogorov-Smirnov test, Pearson chi-square test and Shapiro-Francia test
 \citep[provided by the package \href{http://cran.r-project.org/web/packages/nortest/index.html}{nortest},][]{nortest} and Shapiro-Wilk test
 (provided by the \href{https://stat.ethz.ch/R-manual/R-devel/library/stats/html/shapiro.test.html}{stats} package).
 \item  Astrophysical properties of galaxies and clusters: SDSS magnitudes, g-r colour index and occupancy of clusters.
\end{enumerate}

With this set of features we test different machine learning algorithms such as the Logistic Regression \citep{boot,logreg}, Support Vector Machines \citep{svm,svm1} 
and Random Forest \citep[hereafter RF,][]{breiman01,randomforest}, 
 provided by the R statistical programming language, with the aim of finding merging systems in the complete
 sample of galaxy clusters in our simulated catalogues. 
 
\begin{figure*}
\centering
\includegraphics[width=0.62\textwidth]{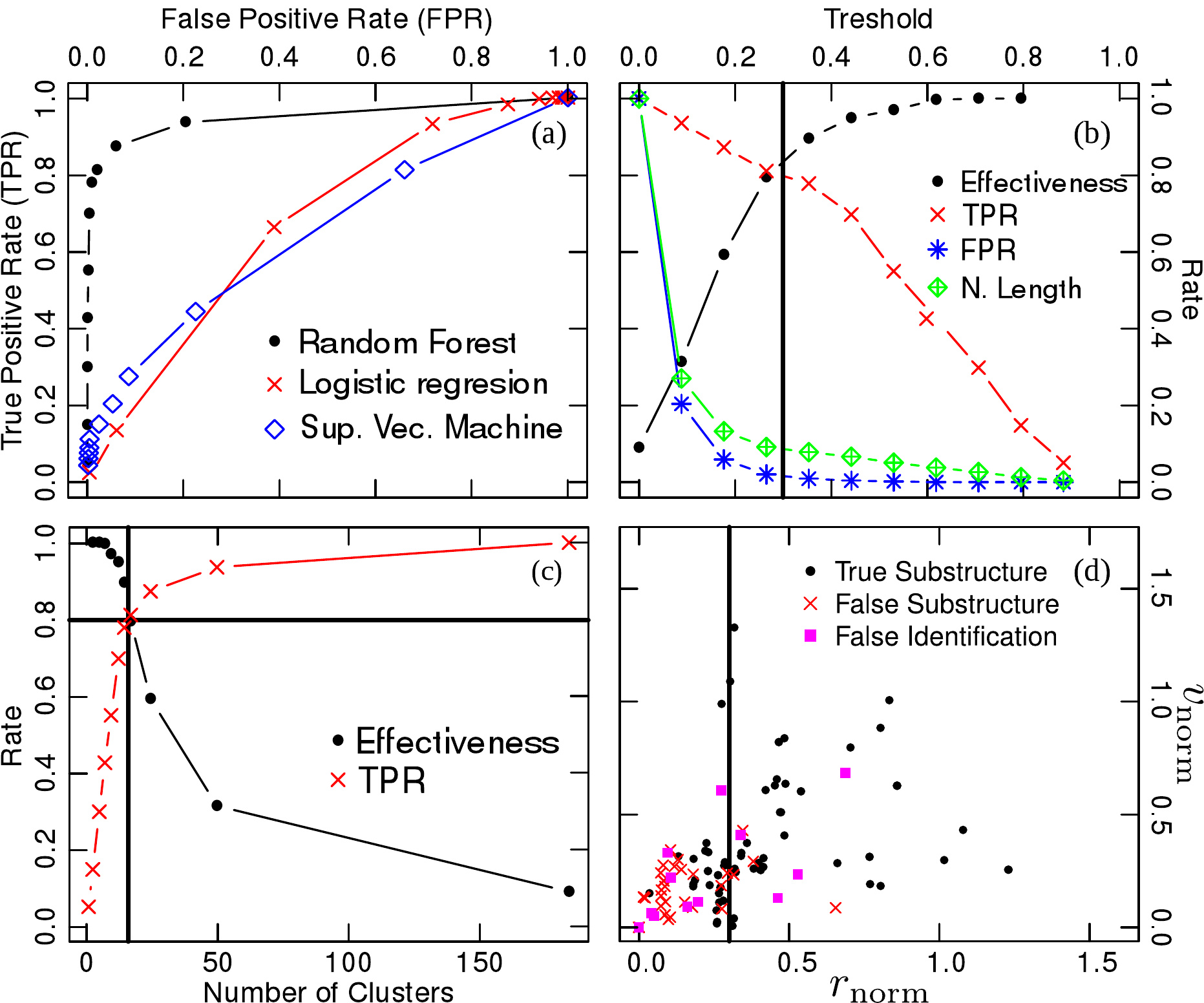}
\caption{\small (a) ROC curve for logistic regression (red crosses), support vector machines (blue diamonds)
 and random forest (black dots) classifiers evaluated using 8 folds cross validation test on the SDSS mock training catalogues. 
(b) Effectiveness (black dots), true positive rate (red crosses), false positive rate (blue asterisks)
 and normalized length (green diamonds) as a function of the threshold imposed to the RF statistic (see text). Vertical line indicates the selected threshold value.
(c) Effectiveness (black dots) and true positive rate (red crosses) as a function of the number of galaxy systems 
for the RF model, with the values at the selected threshold indicated by continuous lines. 
(d) Normalized projected distances $r_\mathrm{norm}$ and velocity difference $v_\mathrm{norm}$ of the merging cluster haloes recovered
 by our methodology (black points). False positive merging clusters are indicated by magenta squares. Merging clusters with missidentified substructures are indicated by red crosses. Vertical line indicates the selection cut introduced in order to avoid LOS contamination.}
\label{fig1}
\end{figure*}

In order to measure its performance we run a standard cross validation test in 8 folds \textit{i.e.}  we divide the total sample in 8 individual and independent subsets and train each machine learning algorithm with 7 of them in order to predict the dynamical status of the clusters of the remaining fold. 
As we know both, the underlying and predicted dynamical status, we are able to compute the true positive rate (TPR), defined as the ratio between the number of merging clusters found in the final sample and the total number of merging clusters that were in the test fold.
In the same way we are able to compute the false positive rate (FPR), defined as the number of relaxed clusters classified as merging clusters divided by the number of relaxed clusters that the studied fold has.
This information allow us to construct the Receiver Operating Characteristic (ROC) curve showed in Fig. \ref{fig1} (a) in which it can be seen that the best performance is obtained by the RF algorithm.

For each cluster the RF computes a statistic that is related to the probability that the cluster is undergoing a merger, by building many decision trees from bootstrap training data, where the final classification
 is based on the average assignation of the ensemble of decision trees. Each tree is grown using randomly selected features from the
 training dataset previously described.
 We impose a threshold to the RF statistic of each cluster in order to classify the sample of merging clusters.
In order to select an optimal threshold, we study in Fig. \ref{fig1} (b) the impact of different values on several statistics of the sample of merging systems, 
namely the TPR, the FPR, the effectiveness (number of identified true--mergers divided by the total number of identified mergers) and the 
normalized length (number of identified mergers divided by the maximum number of identified mergers of the different thresholds). 
As can be seen, the classification threshold impacts on the performance of the RF classifier.
 Consequently, we select a threshold value of $0.3$. This selection criterion guarantees a low false positive detection
 (high effectiveness) in the selected merging systems, but it should be recalled that we are only detecting just a fraction
 of the overall merging systems in the simulated catalogue as can be seen in Fig. \ref{fig1} (c). We note panel (d) of Fig. \ref{fig1} will be analyze later in section 2.3.

The RF implementation also allows us to assess the relative importance of the features as described in \cite{ggrf}. 
In our case the most important features are the number of galaxy members, the p-value of the Shapiro-Wilk test and the Dressler-Shectman test. Nevertheless, it is worth noting that all the previously described features are used in our RF implementation.

\subsection{Measured Properties of the Merging Haloes.} \label{sec:properties}
\begin{figure*}
\centering
\includegraphics[width=0.62\textwidth]{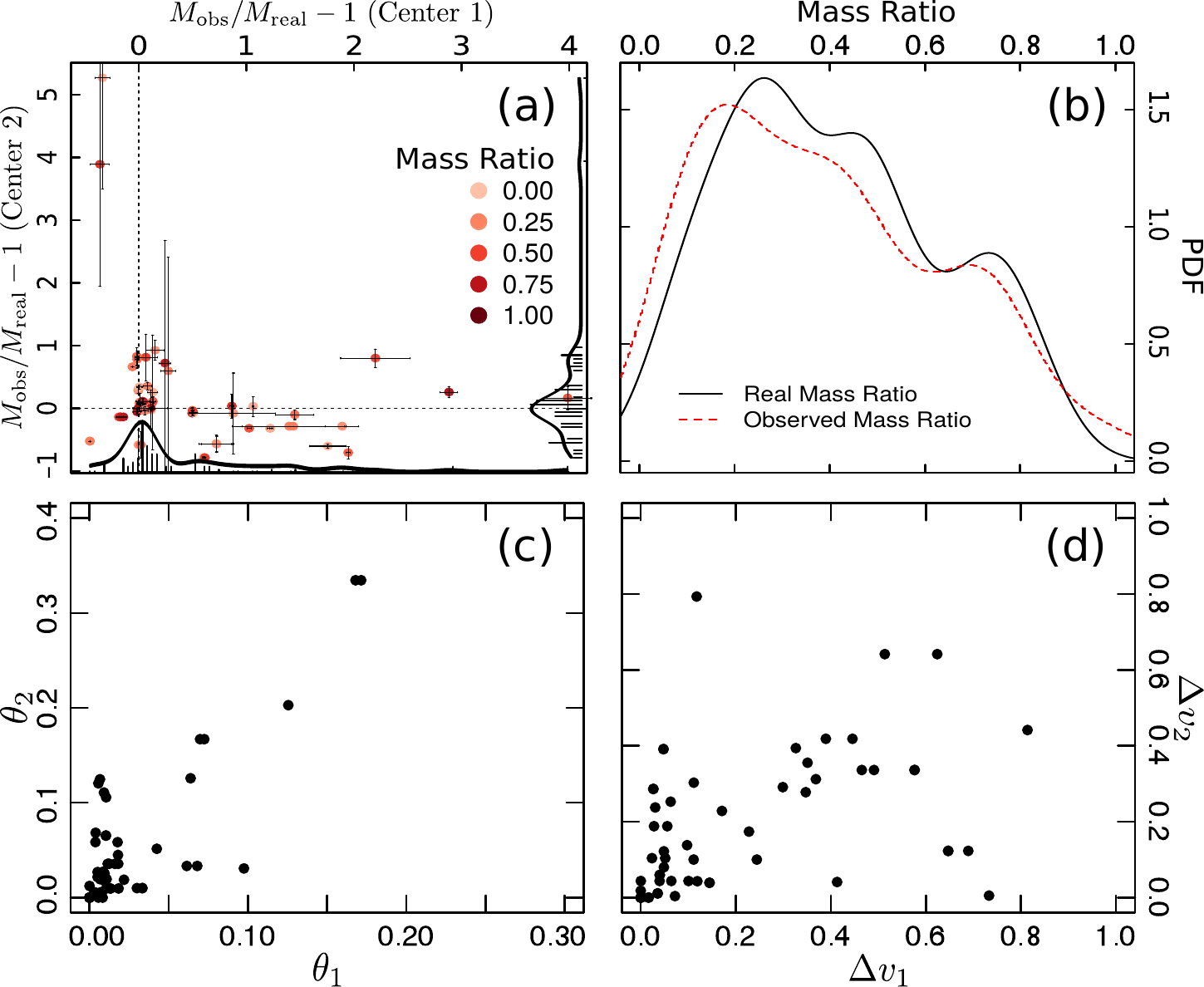}
\caption{\small Recovered properties of merging galaxy systems identified in the simulated mock catalogue, (1) and (2) refers to the main and the merging subhaloes respectively:
(a) Estimated mass bias ($M_\mathrm{obs}/M_\mathrm{real}-1$) for the main and merging substructure. Color scale represent different mass ratios.
(b) Probability distribution function for the real and observed mass ratio ($M_{2}/M_{1}$).
Panel (c) displays the cluster center separation on the sky plane (real and recovered) normalized to the real virial radius.
Panel (d) shows the absolute value of the velocity separation (real and recovered) normalized to the real velocity dispersion. See text for further explanations.}
\label{fig2}
\end{figure*}

The RF algorithm gives us a list of galaxies with high probability of belonging to the merging system. On the other hand, this method does not provide information about galaxy membership to the individual substructures.
Using a Mixture of Gaussian algorithms 
\citep[R package \href{http://cran.r-project.org/web/packages/mclust/index.html}{\textit{mclust},}][]{mclust}
it is possible to cluster these member galaxies into the two merging substructures. Assigning galaxies we are able to compute the centre position
 (angular and redshift) of the substructures, its velocity dispersion
 (Gapper estimator), virial radius and, therefore, measure a dynamical mass. In order to estimate the associated errors we implemented a bootstrap technique. As shown in Fig. \ref{fig2} (a, b) the individual masses and the mass ratios are well recovered.

The recovered geometries of the merging systems can be seen in Fig. \ref{fig1} (d)
in terms of the separation of the components in the line of sight (LOS) and in the plane of the sky. Specifically in the x-axis we compute the normalized projected  distance $r_\mathrm{norm}= d_{1,2}/(r_\mathrm{vir\,1}+r_\mathrm{vir\,2})$, where $d_{1,2}$ is the angular separation between both components of the merging system and $r_\mathrm{vir\,1}$  and $r_\mathrm{vir\,2}$ are the corresponding virial radii. In the y-axis we show 
the velocity difference $v_\mathrm{norm}= |v_{1,2}|/(\sigma_{1}+\sigma_{2})$ where $v_{1,2}$ is the velocity distance between components and $\sigma_{1}$ and $\sigma_{2}$ are the corresponding velocity dispersions of the substructures. 
  Using this identification method we found 3 different cases. Relaxed clusters that we classify as merging clusters (indicated by magenta squares), merging clusters in which we are unable to recover the real substructures (indicated by red crosses) 
  and merging clusters in which we do recover the true substructures that are undergoing a merger (indicated by black dots). As can be seen the 
  false positive cases (FPR $\sim{15}\%$), indicated by magenta squares, are evenly distributed as a function of the $r_{norm}$ parameter. 
  The merging systems where we are not able to recover the real substructures are concentrated below a value of $r_{norm} = 0.22$. 
  Such a selection cut will be used in order to report the properties of the substructures. Although we are able to find spatially coincident 
  merging systems ($r_{norm} < 0.22$, that is systems merging along the LOS or systems with a small projected angular separation) we are not 
  able to recover well the intervening substructures.
  
Using this methodology 
we are able to define samples of merging systems with high levels of
 purity, low contamination and almost exact computation of the centre position of each component, as can be seen in Fig. \ref{fig2} (c,d). In panel (c) we show the angular separation $\theta$ between the actual and the recovered positions of each component,
 normalized to the actual virial radius. Similarly in panel (d)
 we show the differences $\Delta v$ between actual and measured radial velocities, divided by the real velocity dispersion. 
It should be recalled that our machine learning method is able to recover the correct substructures which belong to the major merger as
 identified using the merger tree.
%
\section{Application to low z Clusters.}
\label{sec:samples}
After testing our algorithm on simulated data, we applied our identification algorithms to galaxy systems with more than 30 galaxy members identified in the updated \citep{manuel} catalogue,
based on SDSS-DR7 \citep{sdss} data and two samples of galaxy cluster measurements: WINGS \citep{cava} and HeCS \citep{rines}. 
For the WINGS clusters we compute the g-r colour of the individual galaxies based on the observed b-v colour applying the formulas presented by the \href{http://www.2dfgrs.net/}{2df Collaboration} based on the results of \cite{fukugita}.

We report the following clusters as spatially coincident merging systems candidates: A2593, A2199\checkmark, A2048\checkmark, A3266\checkmark, A3497 (full list will be available in the published version).Notice that we do not provide the substructure properties due to the LOS projection effect (as discussed in section \ref{sec:properties}), therefore minor mergers could be included in this sample \citep{a1201}.
We display in Table \ref{tab1} the properties of the two interacting structures of the merging clusters identified using our
machine learning (RF) classification techniques.
The errors for each property are the standard deviation computed on a hundred of measurements obtained from the RF and the clustering realizations.
Many of them are well known
 merging systems (indicated by \checkmark in the table),
 however it is important to emphasize that our method was able to find several new candidates and also to measure their properties. 
 It should be noted that the decomposition of these structures is only indicative, because galaxy velocities are strongly affected by the gravitational attraction of the two halos.
 Therefore a tomographic reconstruction is necessary (including lensing and X-ray/SZ data) to recover the substructures accurately.
\subsection{The case of multiple major mergers.}

It is well known that there are some clusters that are the result of the merging of more than two systems, although they represent only a small fraction of the total sample (27 of 132 in 8 SDSS mock catalogues).
In order to recover all the merging substructure we performed a mixture of more than two Gaussians. We found that our algorithm is only able to recover, with reliable properties, the two more important substructures. Hence the remaining structures may appear as contamination or not appear at all.
We discuss an individual analysis of Abell 1758 as an example of the multiple major merger case. 
This cluster is known as a merger of four substructures, two in the north and two in the south \citep{a1758}.
At a first iteration, our algorithm, classified correctly Abell 1758 as a merging cluster, but failed to properly reconstruct the merging substructures. 
Considering this extra information we separate the cluster in north and south components and perform a new analysis to each separately.
We find that our algorithm was able to classify both as merging clusters and well recovers the merging substructure properties of both components (north and south).

\section[]{Conclusions and Future works}
\label{sec:conclusions}
In this work we introduce a method aimed to detect merging systems of galaxies in redshift surveys. 
We select a random forest algorithm between other machine learning algorithms, and use as features quantities derived from 
the galaxy redshift space information and from photometry (eg. colours). 
Our detection method was trained and calibrated using a sample of merging systems extracted from mock catalogues.
By studying the merger trees we check that we do find the two substructures that experienced a major merger and recover their fundamental
properties (positions and masses). 
We apply our techniques to a sample of systems of galaxies identified in SDSS-DR7, WINGS and HeCS.

The resulting merging system sample in which we are able to recover the merging substructures comprises 12, 4 and 16 systems respectively. Additionally, we report 29 spatially coincident merging system candidates. 
Several of this systems where previously reported by other authors as interacting systems of galaxies. We also report for the first time 40 new nearby candidates as merging systems, that were previously overlooked.

We emphasize that our method detects in a reliable way the merging systems candidates and substructures properties, but also wish to note that, 
in the case of multiple mergers, some merging substructures may be joined by our algorithm and hence are reported as one component causing a 
possible biase in some measured properties.

\begin{figure*}
\centering
\includegraphics[scale=0.23]{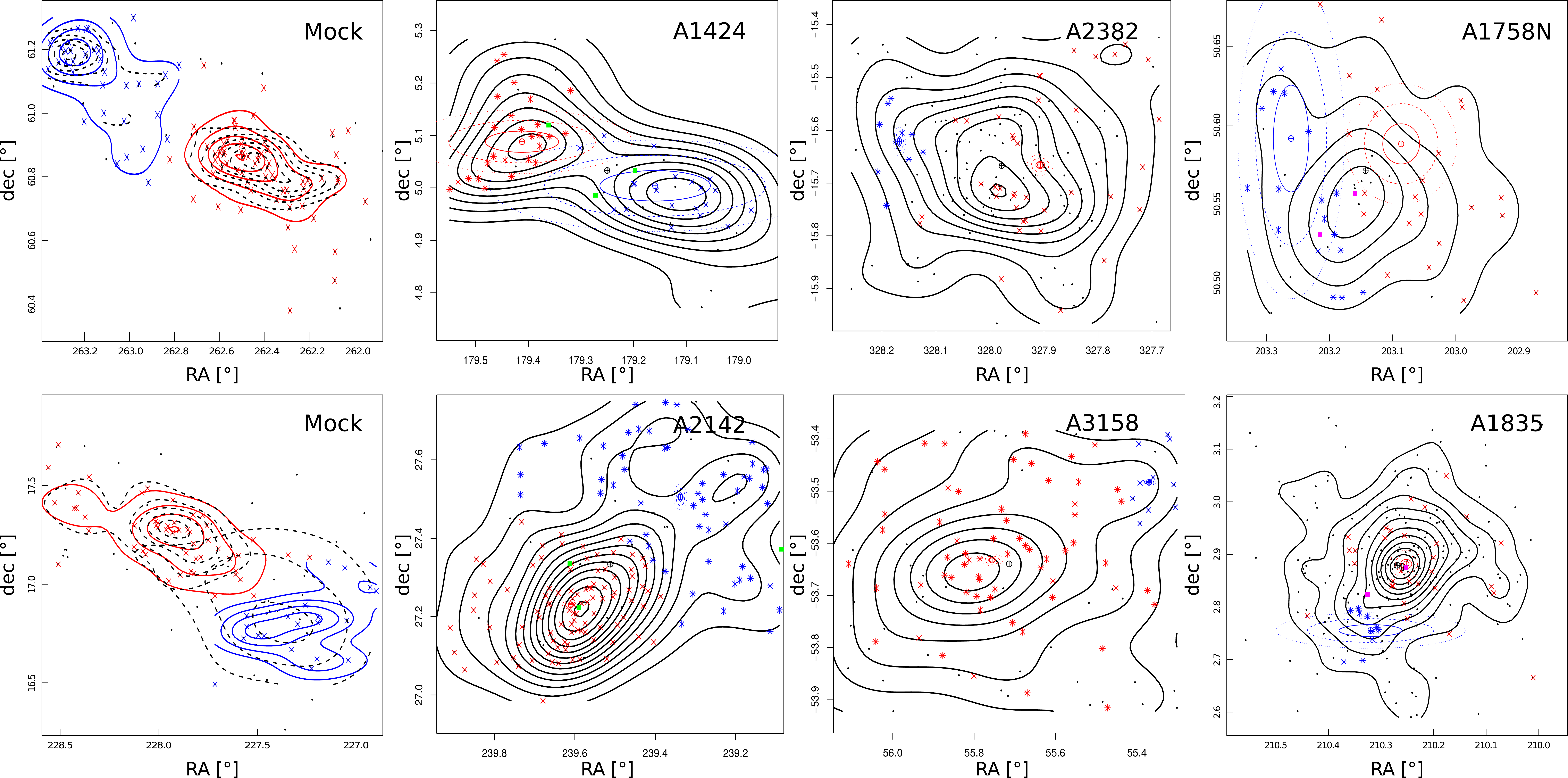}
\caption{\small Galaxy Angular distribution of some merging systems. From left to right we show on columns: two simulated systems, two clusters from the SDSS, two WINGs clusters and two HECs clusters. Member galaxies of both substructures are showed in red
 and blue dots while in black lines are plotted the iso-density contours.
The ellipses indicate the 1, 2 and 3 sigma error of the identified substructure (on blue or red colours depending on the component), with solid, dashed and dotted line, respectively.
For comparison, in green dots are plotted the angular
 positions of the X-ray sources, and in magenta dots are plotted the angular positions of the substructures identified by
 other authors. For the simulated merging systems we also show in dashed line the iso-density of the true substructures that are colliding and in red and blue lines de iso-density contours for the identified substructures.}
\label{fig3}
\end{figure*}

The kinematical reconstructions (see Fig. \ref{fig3} and Table \ref{tab1}) will be corroborated using tomographic techniques including X-ray and weak lensing information \citep{gonzalez} in forthcoming papers.
We also plan in further works to train our algorithm with light cone mocks in order to apply our technique to high redshift catalogues
 (CLASH-VLT \citep{clash}, FRONTIERs \citep{frontier}, EDiCs \citep{edics}, DESI, etc.) and to study any environmental dependence of galaxy properties 
(eg. star formation rate, stellar mass, morphology) at different stages in the merging process.
Diverse studies could be performed with a sample of merging systems like the ones presented in this paper. Using the Bayesian
 reconstruction techniques presented by \cite{2013ApJ...772..131D} it is possible to recover the 3D information of the merger.
In a forthcoming paper we will use such information in order to explore the implications for the properties of the DM particle.
A web interface implementing these methods (The MeSsI Algorithm) is freely available at \href{http://200.16.29.98/martin/merclust}{http://200.16.29.98/martin/merclust}.  

\begin{tiny}
\begin{table*}
\caption{Here we present the low redshift merging cluster sample, including the properties of the two main substructures identified 
by our algorithm. In column 1 we present the name of the cluster, from column 2 to 5 we
 present the estimated mass and the position of the main substructures and from
 column 6 to 9 we present the estimated mass and the position of the other substructure, finally in the last column we list previous work
 on each cluster. Clusters that have been previously reported as merging systems are indicated with \checkmark. 
 References: 
 2 \protect\cite{Wen}, 
 3 \protect\cite{Einasto},
 4 \protect\cite{Cohen},
 6 \protect\cite{Abdullah}, 
 7 \protect\cite{CIRS}, 
 8 \protect\cite{Rhee},
 9 \protect\cite{Parekh},
 16 \protect\cite{Ramella},
 21 \protect\cite{Wang}, 
 22 \protect\cite{Johnston},
 23 \protect\cite{Boschin}, 
 26 \protect\cite{Ragozzine},
 27 \protect\cite{Durret2},
 29 \protect\cite{Smith}, 
 30 \protect\cite{Korngutt}. 
 Full version of the table is available in the online version.}
\begin{center}
\begin{tabular}{llllllllll}

Name  & $M_{1}$ $[10^{14}M_{\odot}]$ & $RA_{1} $[\textdegree] & $DEC_{1} $ [\textdegree] & $z_{1} $ & $M_{2}$ [$10^{14}M_{\odot}$] & $RA_{2} $[\textdegree] & $DEC_{2} $[\textdegree]
& $z_{2} $ & References \\

Abell 1424 & $4.9 $ & $179.38 $ & $ 5.08$ & $0.0760 $ & $5.1$ & $179.19$ & $5.01 $ & $0.0746 $ & 3,6.7,8 \\
 & $\pm 2.3$ & $\pm 0.09$ & $\pm 0.02$ & $\pm 0.0004$ & $\pm 1.4$ & $\pm 0.1$ & $\pm 0.04$ & $\pm 0.0005$ & \\
Abell 2142 & $18.3 $ & $239.61 $ & $27.23 $ & $0.0901 $ & $11.3$ & $239.33$ & $ 27.5$ & $0.0893 $ & 2 \\
\checkmark & $\pm 0.6$ & $\pm 0.005$ & $\pm 0.005$ & $\pm 0.0004$ & $\pm 1.8$ & $\pm 0.005$ & $\pm 0.005$ & $\pm 0.0001$ & \\
Abell 3158  & $37.24 $   & $55.75 $   & $-53.63 $   & $0.0633 $    & $4.6$      & $55.37 $     & $-53.48 $   & $0.0622 $ & 4,9,16,21,22 \\
\checkmark  & $\pm 1.5$  & $\pm 0.07$ & $\pm 0.004$ & $\pm 0.0001$ & $ \pm 0.2$ & $\pm 0.007 $ & $\pm 0.001$ & $\pm 0.0001$ &  \\
Abell 2382 & $77.7$     & $327.90 $   & $-15.66 $   & $0.0676 $    & $6.12$     & $328.167 $   & $-15.62 $  & $0.0642$     & \\
            & $\pm 10.2$ & $\pm 0.006$ & $\pm 0.006$ & $\pm 0.0003$ & $ \pm 1.1$ & $\pm 0.003$  & $\pm 0.01$ & $\pm 0.0002$ &  \\
Abell 1758N & $59.3 $ & $203.07 $ & $50.59 $ & $0.2768 $ & $29.1$ & $203.25$ & $50.57 $ & $0.2783 $ & 2,23,26,27 \\
\checkmark & $\pm 9$ & $\pm 0.02$ & $\pm 0.01$ & $\pm 0.0002$ & $\pm 15.8$ & $\pm 0.02$ & $\pm 0.03$ & $\pm 0.0007$ & \\
Abell 1835 & $ 105$ & $ 210.25$ & $2.87 $ & $0.2516 $ & $17$ & $210.29$ & $ 2.75$ & $0.2479 $ & 2,29,30 \\
 & $\pm 14$ & $\pm 0.01$ & $\pm 0.01$ & $\pm 0.0006$ & $\pm 25$ & $\pm 0.09$ & $\pm 0.04$ & $\pm 0.002$ & \\

\label{tab1}
\end{tabular}
\end{center}
\end{table*}
\end{tiny}

\section*{Acknowledgements}

MdlR warmly thanks to Monica Jauregui and Jorge de los Rios for the unconditional support.
This work is dedicated with admiration to Lionel Messi. 
We also thank to Dario Gra\~na for the technical support and Sebast\'ian Gurovich for useful comments.
This work was partially supported by the Consejo Nacional de Investigaciones 
Cient\'{\i}ficas y T\'ecnicas (CONICET, Argentina) 
and the Secretar\'{\i}a de Ciencia y Tecnolog\'{\i}a de la Universidad Nacional 
de C\'ordoba (SeCyT-UNC, Argentina).
This research was developed while MJdLDR was a postdoctoral researcher at the 
\href{Center for Gravitational Wave Astronomy at
 The University of Texas in Brownsville}{http://cgwa.phys.utb.edu/}.
This research has made use of the
\href{NASA's Astrophysics Data System}{http://adsabs.harvard.edu/}, Cornell University \href{arXiv}{xxx.arxiv.org} repository, the SIMBAD database,
operated at CDS, Strasbourg, France, the \href{http://www.r-project.org/}{R project for statistical computing} and the following
 \href{http://mirror.fcaglp.unlp.edu.ar/CRAN/}{CRAN} libraries: \texttt{mclust,rgl,randomForest,nortest,e1071,shiny}.\\
The Millennium Simulation databases \cite{lemson1} used in this paper and the web application providing online access to
 them were constructed as part of the activities of the GAVO. \\
Funding for the SDSS and SDSS-II has been provided by the Alfred P. Sloan Foundation, the Participating Institutions, 
the National Science Foundation, the U.S. Department of Energy, the National Aeronautics and Space Administration, the 
Japanese Monbukagakusho, the Max Planck Society, and the Higher Education Funding Council for England. 
The SDSS Web Site is 
\href{http://www.sdss.org/}{http://www.sdss.org/}.
\bibliographystyle{mn2e}
\bibliography{references}

\begin{thebibliography}{}

\bibitem[\protect\citeauthoryear{{Abazajian} et~al.,}{{Abazajian}
  et~al.}{2009}]{sdss}
{Abazajian} K.~N.,  et~al., 2009, \apjs, 182, 543

\bibitem[\protect\citeauthoryear{{Abdullah}, {Ali}, {Ismail} \&
  {Rassem}}{{Abdullah} et~al.}{2011}]{Abdullah}
{Abdullah} M.~H.,  {Ali} G.~B.,  {Ismail} H.~A.,    {Rassem} M.~A.,  2011,
  \mnras, 416, 2027

\bibitem[\protect\citeauthoryear{{Andrade-Santos} et~al.,}{{Andrade-Santos}
  et~al.}{2015}]{andrade}
{Andrade-Santos} F.,  et~al., 2015, ArXiv e-prints

\bibitem[\protect\citeauthoryear{{Biviano}, {Rosati}, {Balestra} \&
  {Mercurio}}{{Biviano} et~al.}{2013}]{clash}
{Biviano} A.,  {Rosati} P.,  {Balestra} I.,    {Mercurio} A.,  2013, \aap, 558,
  A1

\bibitem[\protect\citeauthoryear{{Boschin}, {Girardi}, {Barrena} \&
  {Nonino}}{{Boschin} et~al.}{2012}]{Boschin}
{Boschin} W.,  {Girardi} M.,  {Barrena} R.,    {Nonino} M.,  2012, \aap, 540,
  A43

\bibitem[\protect\citeauthoryear{{Bradac}, {Allen}, {Treu}, {Ebeling},
  {Massey}, {Morris}, {von der Linden} \& {Applegate}}{{Bradac}
  et~al.}{2008}]{bradac}
{Bradac} M.,  {Allen} S.~W.,  {Treu} T.,  {Ebeling} H.,  {Massey} R.,  {Morris}
  R.~G.,  {von der Linden} A.,    {Applegate} D.,  2008, \apj, 687, 959

\bibitem[\protect\citeauthoryear{Breiman}{Breiman}{2001}]{breiman01}
Breiman L.,  2001, Machine Learning, 45, 5

\bibitem[\protect\citeauthoryear{Canty \& Ripley}{Canty \&
  Ripley}{2015}]{logreg}
Canty A.,  Ripley B.~D.,  2015, boot: Bootstrap R Functions

\bibitem[\protect\citeauthoryear{{Cava} et~al.,}{{Cava}  et~al.}{2009}]{cava}
{Cava} A.,  et~al., 2009, \aap, 495, 707

\bibitem[\protect\citeauthoryear{{Clowe}, {Bradac}, {Gonzalez}, {Markevitch},
  {Randall}, {Jones} \& {Zaritsky}}{{Clowe} et~al.}{2006}]{clowe}
{Clowe} D.,  {Bradac} M.,  {Gonzalez} A.~H.,  {Markevitch} M.,  {Randall}
  S.~W.,  {Jones} C.,    {Zaritsky} D.,  2006, \apjl, 648, L109

\bibitem[\protect\citeauthoryear{{Clowe}, {Markevitch}, {Bradac}, {Gonzalez},
  {Chung}, {Massey} \& {Zaritsky}}{{Clowe} et~al.}{2012}]{clowe2012}
{Clowe} D.,  {Markevitch} M.,  {Bradac} M.,  {Gonzalez} A.~H.,  {Chung} S.~M.,
  {Massey} R.,    {Zaritsky} D.,  2012, \apj, 758, 128

\bibitem[\protect\citeauthoryear{{Cohen}, {Hickox}, {Wegner}, {Einasto} \&
  {Vennik}}{{Cohen} et~al.}{2014}]{Cohen}
{Cohen} S.~A.,  {Hickox} R.~C.,  {Wegner} G.~A.,  {Einasto} M.,    {Vennik} J.,
   2014, \apj, 783, 136

\bibitem[\protect\citeauthoryear{Cortes \& Vapnik}{Cortes \&
  Vapnik}{1995}]{svm}
Cortes C.,  Vapnik V.,  1995, Machine Learning, 20, 273

\bibitem[\protect\citeauthoryear{Davison \& Hinkley}{Davison \&
  Hinkley}{1997}]{boot}
Davison A.~C.,  Hinkley D.~V.,  1997, Bootstrap Methods and Their Applications.
CUP, Cambridge

\bibitem[\protect\citeauthoryear{{Dawson}}{{Dawson}}{2013}]{2013ApJ...772..131D}
{Dawson} W.~A.,  2013, \apj, 772, 131

\bibitem[\protect\citeauthoryear{{Dawson} et~al.,}{{Dawson}
  et~al.}{2012}]{dawson}
{Dawson} W.~A.,  et~al., 2012, \apjl, 747, L42

\bibitem[\protect\citeauthoryear{{Dom{\'{\i}}nguez Romero}, {Garc{\'{\i}}a
  Lambas} \& {Muriel}}{{Dom{\'{\i}}nguez Romero} et~al.}{2012}]{dominguez}
{Dom{\'{\i}}nguez Romero} M.,  {Garc{\'{\i}}a Lambas} D.,    {Muriel} H.,
  2012, \mnras, 427, L6

\bibitem[\protect\citeauthoryear{{Dressler} \& {Shectman}}{{Dressler} \&
  {Shectman}}{1988}]{dressler}
{Dressler} A.,  {Shectman} S.~A.,  1988, \aj, 95, 985

\bibitem[\protect\citeauthoryear{{Durret}, {Lagan{\'a}} \& {Haider}}{{Durret}
  et~al.}{2011}]{Durret2}
{Durret} F.,  {Lagan{\'a}} T.~F.,    {Haider} M.,  2011, \aap, 529, A38

\bibitem[\protect\citeauthoryear{{Durret}, {Lima Neto} \& {Forman}}{{Durret}
  et~al.}{2005}]{durret}
{Durret} F.,  {Lima Neto} G.,    {Forman} W.,  2005, \aap, 432, 809

\bibitem[\protect\citeauthoryear{{Ebeling}, {Ma} \& {Barrett}}{{Ebeling}
  et~al.}{2014}]{frontier}
{Ebeling} H.,  {Ma} C.-J.,    {Barrett} E.,  2014, \apjs, 211, 21

\bibitem[\protect\citeauthoryear{Ehrlinger}{Ehrlinger}{2015}]{ggrf}
Ehrlinger J.,  2015, ggRandomForests: Graphical Exploration of Random Forests

\bibitem[\protect\citeauthoryear{{Einasto} et~al.,}{{Einasto}
  et~al.}{2012}]{Einasto}
{Einasto} M.,  et~al., 2012, \aap, 540, A123

\bibitem[\protect\citeauthoryear{{Farrar} \& {Rosen}}{{Farrar} \&
  {Rosen}}{2007}]{farrar}
{Farrar} G.~R.,  {Rosen} R.~A.,  2007, Physical Review Letters, 98, 171302

\bibitem[\protect\citeauthoryear{{Feretti}, {Giovannini}, {Govoni} \&
  {Murgia}}{{Feretti} et~al.}{2012}]{feretti}
{Feretti} L.,  {Giovannini} G.,  {Govoni} F.,    {Murgia} M.,  2012, \aapr, 20,
  54

\bibitem[\protect\citeauthoryear{{Forero-Romero}, {Gottlober} \&
  {Yepes}}{{Forero-Romero} et~al.}{2010}]{forero}
{Forero-Romero} J.~E.,  {Gottlober} S.,    {Yepes} G.,  2010, \apj, 725, 598

\bibitem[\protect\citeauthoryear{Fraley, Raftery, Murphy \& Scrucca}{Fraley
  et~al.}{2012}]{mclust}
Fraley C.,  Raftery A.~E.,  Murphy T.~B.,    Scrucca L.,  2012

\bibitem[\protect\citeauthoryear{{Fukugita}, {Shimasaku} \&
  {Ichikawa}}{{Fukugita} et~al.}{1995}]{fukugita}
{Fukugita} M.,  {Shimasaku} K.,    {Ichikawa} T.,  1995, pasp, 107, 945

\bibitem[\protect\citeauthoryear{{Gastaldello} et~al.,}{{Gastaldello}
  et~al.}{2014}]{gael}
{Gastaldello} F.,  et~al., 2014, \mnras, 442, L76

\bibitem[\protect\citeauthoryear{{Gonzalez} et~al.,}{{Gonzalez}
  et~al.}{2015}]{gonzalez}
{Gonzalez} J.~E.,  et~al., 2015, ArXiv e-prints

\bibitem[\protect\citeauthoryear{Gross \& Ligges}{Gross \&
  Ligges}{2015}]{nortest}
Gross J.,  Ligges U.,  2015, nortest: Tests for Normality

\bibitem[\protect\citeauthoryear{{Guo} et~al.,}{{Guo}  et~al.}{2011}]{guo}
{Guo} Q.,  et~al., 2011, \mnras, 413, 101

\bibitem[\protect\citeauthoryear{{Harvey} et~al.,}{{Harvey}
  et~al.}{2014}]{harvey}
{Harvey} D.,  et~al., 2014, \mnras, 441, 404

\bibitem[\protect\citeauthoryear{{Harvey}, {Massey}, {Kitching}, {Taylor} \&
  {Tittley}}{{Harvey} et~al.}{2015}]{harveytwo}
{Harvey} D.,  {Massey} R.,  {Kitching} T.,  {Taylor} A.,    {Tittley} E.,
  2015, ArXiv e-prints

\bibitem[\protect\citeauthoryear{{Hayashi} \& {White}}{{Hayashi} \&
  {White}}{2006}]{hayashi}
{Hayashi} E.,  {White} S.~D.~M.,  2006, \mnras, 370, L38

\bibitem[\protect\citeauthoryear{{Hou}, {Parker} \& {Harris}}{{Hou}
  et~al.}{2014}]{hou}
{Hou} A.,  {Parker} L.,    {Harris} W.~E.,  2014, \mnras, 442, 406

\bibitem[\protect\citeauthoryear{{Jee}, {Hoekstra}, {Mahdavi} \& {Babul}}{{Jee}
  et~al.}{2014}]{jee2014}
{Jee} M.~J.,  {Hoekstra} H.,  {Mahdavi} A.,    {Babul} A.,  2014, ArXiv
  e-prints

\bibitem[\protect\citeauthoryear{{Jee}, {Mahdavi}, {Hoekstra}, {Babul},
  {Dalcanton}, {Carroll} \& {Capak}}{{Jee} et~al.}{2012}]{jee2012}
{Jee} M.~J.,  {Mahdavi} A.,  {Hoekstra} H.,  {Babul} A.,  {Dalcanton} J.~J.,
  {Carroll} P.,    {Capak} P.,  2012, \apj, 747, 96

\bibitem[\protect\citeauthoryear{{Johnston-Hollitt}, {Sato}, {Gill}, {Fleenor}
  \& {Brick}}{{Johnston-Hollitt} et~al.}{2008}]{Johnston}
{Johnston-Hollitt} M.,  {Sato} M.,  {Gill} J.~A.,  {Fleenor} M.~C.,    {Brick}
  A.-M.,  2008, \mnras, 390, 289

\bibitem[\protect\citeauthoryear{{Kahlhoefer}, {Schmidt-Hoberg}, {Frandsen} \&
  {Sarkar}}{{Kahlhoefer} et~al.}{2014}]{kahlhoefer}
{Kahlhoefer} F.,  {Schmidt-Hoberg} K.,  {Frandsen} M.~T.,    {Sarkar} S.,
  2014, \mnras, 437, 2865

\bibitem[\protect\citeauthoryear{{Korngut} et~al.,}{{Korngut}
  et~al.}{2011}]{Korngutt}
{Korngut} P.~M.,  et~al., 2011, \apj, 734, 10

\bibitem[\protect\citeauthoryear{{Lares}, {Lambas} \&
  {Dom{\'{\i}}nguez}}{{Lares} et~al.}{2011}]{lares}
{Lares} M.,  {Lambas} D.,    {Dom{\'{\i}}nguez} M.,  2011, \aj, 142, 13

\bibitem[\protect\citeauthoryear{{Lee} \& {Komatsu}}{{Lee} \&
  {Komatsu}}{2010}]{lee2}
{Lee} J.,  {Komatsu} E.,  2010, \apj, 718, 60

\bibitem[\protect\citeauthoryear{{Lemson} \& {Virgo Consortium}}{{Lemson} \&
  {Virgo Consortium}}{2006}]{lemson1}
{Lemson} G.,  {Virgo Consortium} t.,  2006, ArXiv Astrophysics e-prints

\bibitem[\protect\citeauthoryear{Liaw \& Wiener}{Liaw \&
  Wiener}{2002}]{randomforest}
Liaw A.,  Wiener M.,  2002, R News, 2, 18

\bibitem[\protect\citeauthoryear{{Ma}, {Owers}, {Nulsen}, {McNamara}, {Murray}
  \& {Couch}}{{Ma} et~al.}{2012}]{a1201}
{Ma} C.-J.,  {Owers} M.,  {Nulsen} P.~E.~J.,  {McNamara} B.~R.,  {Murray}
  S.~S.,    {Couch} W.~J.,  2012, \apj, 752, 139

\bibitem[\protect\citeauthoryear{{Mahdavi}, {Hoekstra}, {Babul}, {Balam} \&
  {Capak}}{{Mahdavi} et~al.}{2007}]{mahdavi}
{Mahdavi} A.,  {Hoekstra} H.,  {Babul} A.,  {Balam} D.~D.,    {Capak} P.~L.,
  2007, \apj, 668, 806

\bibitem[\protect\citeauthoryear{{Mann} \& {Ebeling}}{{Mann} \&
  {Ebeling}}{2012}]{mann}
{Mann} A.~W.,  {Ebeling} H.,  2012, \mnras, 420, 2120

\bibitem[\protect\citeauthoryear{{Markevitch}, {Gonzalez}, {Clowe},
  {Vikhlinin}, {Forman}, {Jones}, {Murray} \& {Tucker}}{{Markevitch}
  et~al.}{2004}]{markevitch}
{Markevitch} M.,  {Gonzalez} A.~H.,  {Clowe} D.,  {Vikhlinin} A.,  {Forman} W.,
   {Jones} C.,  {Murray} S.,    {Tucker} W.,  2004, \apj, 606, 819

\bibitem[\protect\citeauthoryear{{Massey}, {Kitching} \& {Nagai}}{{Massey}
  et~al.}{2011}]{massey}
{Massey} R.,  {Kitching} T.,    {Nagai} D.,  2011, \mnras, 413, 1709

\bibitem[\protect\citeauthoryear{{Mastropietro} \& {Burkert}}{{Mastropietro} \&
  {Burkert}}{2008}]{mastro}
{Mastropietro} C.,  {Burkert} A.,  2008, \mnras, 389, 967

\bibitem[\protect\citeauthoryear{{Menanteau} et~al.,}{{Menanteau}
  et~al.}{2012}]{menanteau}
{Menanteau} F.,  et~al., 2012, \apj, 748, 7

\bibitem[\protect\citeauthoryear{{Merch{\'a}n} \& {Zandivarez}}{{Merch{\'a}n}
  \& {Zandivarez}}{2002}]{manuel}
{Merch{\'a}n} M.,  {Zandivarez} A.,  2002, \mnras, 335, 216

\bibitem[\protect\citeauthoryear{{Merch{\'a}n} \& {Zandivarez}}{{Merch{\'a}n}
  \& {Zandivarez}}{2005}]{ariel}
{Merch{\'a}n} M.~E.,  {Zandivarez} A.,  2005, \apj, 630, 759

\bibitem[\protect\citeauthoryear{{Merten} et~al.,}{{Merten}
  et~al.}{2011}]{merten}
{Merten} J.,  et~al., 2011, \mnras, 417, 333

\bibitem[\protect\citeauthoryear{Meyer, Dimitriadou, Hornik, Weingessel \&
  Leisch}{Meyer et~al.}{2014}]{svm1}
Meyer D.,  Dimitriadou E.,  Hornik K.,  Weingessel A.,    Leisch F.,  2014,
  e1071, TU Wien

\bibitem[\protect\citeauthoryear{{Milosavljevic}, {Koda}, {Nagai}, {Nakar} \&
  {Shapiro}}{{Milosavljevic} et~al.}{2007}]{milosav}
{Milosavljevic} M.,  {Koda} J.,  {Nagai} D.,  {Nakar} E.,    {Shapiro} P.~R.,
  2007, \apjl, 661, L131

\bibitem[\protect\citeauthoryear{{Milvang-Jensen}, {Noll}, {Halliday} \&
  {Poggianti}}{{Milvang-Jensen} et~al.}{2008}]{edics}
{Milvang-Jensen} B.,  {Noll} S.,  {Halliday} C.,    {Poggianti} B.~M.,  2008,
  \aap, 482, 419

\bibitem[\protect\citeauthoryear{{Mo}, {Van den Bosch} \& {White}}{{Mo}
  et~al.}{2010}]{Mo_Van_den_Bosch_White_2010}
{Mo} H.,  {Van den Bosch} F.,    {White} S.,  2010, Galaxy formation and
  evolution.
CUP

\bibitem[\protect\citeauthoryear{{Molnar} \& {Broadhurst}}{{Molnar} \&
  {Broadhurst}}{2015}]{molnar}
{Molnar} S.~M.,  {Broadhurst} T.,  2015, \apj, 800, 37

\bibitem[\protect\citeauthoryear{{Ng}, {Dawson}, {Wittman}, {Jee}, {Hughes},
  {Menanteau} \& {Sif{\'o}n}}{{Ng} et~al.}{2014}]{dawsongordo}
{Ng} K.~Y.,  {Dawson} W.~A.,  {Wittman} D.,  {Jee} M.~J.,  {Hughes} J.~P.,
  {Menanteau} F.,    {Sif{\'o}n} C.,  2014, ArXiv e-prints

\bibitem[\protect\citeauthoryear{{Parekh}, {van der Heyden}, {Ferrari}, {Angus}
  \& {Holwerda}}{{Parekh} et~al.}{2015}]{Parekh}
{Parekh} V.,  {van der Heyden} K.,  {Ferrari} C.,  {Angus} G.,    {Holwerda}
  B.,  2015, \aap, 575, A127

\bibitem[\protect\citeauthoryear{{Pinkney}, {Roettiger}, {Burns} \&
  {Bird}}{{Pinkney} et~al.}{1996}]{pinkney}
{Pinkney} J.,  {Roettiger} K.,  {Burns} J.~O.,    {Bird} C.~M.,  1996, \apjs,
  104, 1

\bibitem[\protect\citeauthoryear{{Ragozzine}, {Clowe}, {Markevitch}, {Gonzalez}
  \& {Brada{\v c}}}{{Ragozzine} et~al.}{2012a}]{a1758}
{Ragozzine} B.,  {Clowe} D.,  {Markevitch} M.,  {Gonzalez} A.~H.,    {Brada{\v
  c}} M.,  2012a, \apj, 744, 94

\bibitem[\protect\citeauthoryear{{Ragozzine}, {Clowe}, {Markevitch}, {Gonzalez}
  \& {Brada{\v c}}}{{Ragozzine} et~al.}{2012b}]{Ragozzine}
{Ragozzine} B.,  {Clowe} D.,  {Markevitch} M.,  {Gonzalez} A.~H.,    {Brada{\v
  c}} M.,  2012b, \apj, 744, 94

\bibitem[\protect\citeauthoryear{{Ramella} et~al.,}{{Ramella}
  et~al.}{2007}]{Ramella}
{Ramella} M.,  et~al., 2007, \aap, 470, 39

\bibitem[\protect\citeauthoryear{{Randall}, {Markevitch}, {Clowe}, {Gonzalez}
  \& {Bradac}}{{Randall} et~al.}{2008}]{randall}
{Randall} S.~W.,  {Markevitch} M.,  {Clowe} D.,  {Gonzalez} A.~H.,    {Bradac}
  M.,  2008, \apj, 679, 1173

\bibitem[\protect\citeauthoryear{{Rhee}, {van Haarlem} \& {Katgert}}{{Rhee}
  et~al.}{1991}]{Rhee}
{Rhee} G.~F.~R.~N.,  {van Haarlem} M.~P.,    {Katgert} P.,  1991, \aap, 246,
  301

\bibitem[\protect\citeauthoryear{{Rines} \& {Diaferio}}{{Rines} \&
  {Diaferio}}{2006}]{CIRS}
{Rines} K.,  {Diaferio} A.,  2006, \aj, 132, 1275

\bibitem[\protect\citeauthoryear{{Rines}, {Geller}, {Diaferio} \&
  {Kurtz}}{{Rines} et~al.}{2013}]{rines}
{Rines} K.,  {Geller} M.~J.,  {Diaferio} A.,    {Kurtz} M.~J.,  2013, \apj,
  767, 15

\bibitem[\protect\citeauthoryear{{Roukema}, {Quinn} \& {Peterson}}{{Roukema}
  et~al.}{1993}]{roukema}
{Roukema} B.~F.,  {Quinn} P.~J.,    {Peterson} B.~A.,  1993, in {Chincarini}
  G.~L.,  {Iovino} A.,  {Maccacaro} T.,   {Maccagni} D.,  eds, Observational
  Cosmology Vol.~51 of Astronomical Society of the Pacific Conference Series,
  {Spectral Evolution of Merging/Accreting Galaxies}.
p.~298

\bibitem[\protect\citeauthoryear{{Smith} \& {Taylor}}{{Smith} \&
  {Taylor}}{2008}]{Smith}
{Smith} G.~P.,  {Taylor} J.~E.,  2008, \apjl, 682, L73

\bibitem[\protect\citeauthoryear{{Solanes}, {Salvador-Sol{\'e}} \&
  {Gonz{\'a}lez-Casado}}{{Solanes} et~al.}{1999}]{solanes}
{Solanes} J.~M.,  {Salvador-Sol{\'e}} E.,    {Gonz{\'a}lez-Casado} G.,  1999,
  \aap, 343, 733

\bibitem[\protect\citeauthoryear{{Springel} et~al.,}{{Springel}
  et~al.}{2005}]{springel:05}
{Springel} V.,  et~al., 2005, \nat, 435, 629

\bibitem[\protect\citeauthoryear{{Springel} \& {Farrar}}{{Springel} \&
  {Farrar}}{2007}]{springel}
{Springel} V.,  {Farrar} G.~R.,  2007, \mnras, 380, 911

\bibitem[\protect\citeauthoryear{{Thompson}, {Dav{\'e}} \&
  {Nagamine}}{{Thompson} et~al.}{2014}]{challenger}
{Thompson} R.,  {Dav{\'e}} R.,    {Nagamine} K.,  2014, ArXiv e-prints

\bibitem[\protect\citeauthoryear{{Thompson} \& {Nagamine}}{{Thompson} \&
  {Nagamine}}{2012}]{thompson}
{Thompson} R.,  {Nagamine} K.,  2012, \mnras, 419, 3560

\bibitem[\protect\citeauthoryear{{Wang}, {Xu}, {Gu}, {Gu}, {Qin}, {Wang},
  {Zhang} \& {Wu}}{{Wang} et~al.}{2010}]{Wang}
{Wang} Y.,  {Xu} H.,  {Gu} L.,  {Gu} J.,  {Qin} Z.,  {Wang} J.,  {Zhang} Z.,
  {Wu} X.-P.,  2010, \mnras, 403, 1909

\bibitem[\protect\citeauthoryear{{Watson}, {Iliev}, {Diego}, {Gottlober},
  {Knebe}, {Martinez-Gonzalez} \& {Yepes}}{{Watson} et~al.}{2014}]{watson}
{Watson} W.~A.,  {Iliev} I.,  {Diego} J.,  {Gottlober} S.,  {Knebe} A.,
  {Martinez-Gonzalez} E.,    {Yepes} G.,  2014, \mnras, 437, 3776

\bibitem[\protect\citeauthoryear{{Wen} \& {Han}}{{Wen} \& {Han}}{2013}]{Wen}
{Wen} Z.~L.,  {Han} J.~L.,  2013, \mnras, 436, 275

\bibitem[\protect\citeauthoryear{{Yu}, {Serra}, {Diaferio} \& {Baldi}}{{Yu}
  et~al.}{2015}]{serra}
{Yu} H.,  {Serra} A.,  {Diaferio} A.,    {Baldi} M.,  2015, ArXiv e-prints

\end{thebibliography}

\label{lastpage}

\end{document}